\documentclass[12pt]{article}

\usepackage{newtxtext,newtxmath}

\usepackage{graphicx}

\usepackage[letterpaper,margin=1in]{geometry}

\renewenvironment{abstract}
	{\quotation}
	{\endquotation}

\date{}

\makeatletter
\renewcommand{\fnum@figure}{\textbf{Figure \thefigure}}
\renewcommand{\fnum@table}{\textbf{Table \thetable}}
\makeatother

\usepackage{scicite}

\usepackage{url}
\usepackage{soul}

\def\scititle{
 Visible to Longwave-infrared imaging via an inverse-designed monolithic lens
}
\title{\bfseries \boldmath \scititle}

\author{
	Syed N. Qadri$^{1\dagger}$,
	Apratim Majumder$^{2\dagger}$,
    John D. Hodges$^{1}$,
	Nicole Brimhall$^{3}$,\\
    Freddie Santiago$^{1}$,
    Rajesh Menon$^{2,3\ast}$\and
	\small$^{1}$U.S. Naval Research Laboratory Remote Sensing Division, Washington DC 20375, USA.\and
	\small$^{2}$Dept. of Electrical \& Computer Engineering, University of Utah, Salt Lake City UT 84112, USA.\and
    \small$^{3}$Oblate Optics, Inc., San Diego, CA 92130, USA.\and
	\small$^\ast$Corresponding author: rmenon@eng.utah.edu \and
	\small$^\dagger$These authors contributed equally to this work.
}

\begin{document} 

\maketitle

\begin{abstract} \bfseries \boldmath
Chromatic aberrations impose a fundamental barrier on optical design, confining most imaging systems to narrow spectral bands with fractional bandwidths typically limited to $\Delta\lambda/\lambda < 1$. Here we report a monolithic, inverse-designed potassium bromide (KBr) lens that achieves broadband, near-achromatic focusing from 0.45 to 14 $\mu$m, a continuous spectral span covering the visible, near-, mid-, and long-wave infrared. This corresponds to a fractional bandwidth of 1.9, approaching the theoretical limit of 2, while maintaining a nearly constant focal length across the entire range. The 19-mm-diameter, 22.5-mm-focal-length optic enables a single compact platform for hyperspectral imaging, mid-IR microscopy, super-resolution, imaging through scattering media, and simultaneous multi-band and long-range imaging. Coupling the KBr lens with a conventional refractive element further yields a hybrid telescope that extends these capabilities. By uniting inverse design with scalable manufacturing, this approach provides a route toward broadly deployable ultra-broadband imagers for biomedicine, climate and environmental monitoring, and space-based sensing.
\end{abstract}

\noindent

Broadband imaging systems that operate across wide spectral ranges underpin major advances in science and technology. In astrophysics, broad spectral access reveals phenomena across cosmic time, from redshifted light of the earliest galaxies to molecular absorption features in exoplanet atmospheres. The James Webb Space Telescope spans 0.6 to 28.5 $\mu$m using a highly complex reflective architecture built from segmented, gold-coated beryllium mirrors \cite{gardner2006james}. Reflective systems avoid chromatic aberration, but require extreme alignment precision, restrict optical geometry, and impose substantial mass and cost. A transmissive platform with comparably broad achromatic bandwidth would simplify instrument design while expanding access to multi-octave imaging. The impact would extend across fields: in remote sensing, simultaneous VIS–NIR–MWIR–LWIR imaging improves material identification and atmospheric analysis \cite{goetz1985imaging,clark2024imaging}; in biomedicine, broad spectral contrast enables label-free diagnostics \cite{yoon2022hyperspectral,bouma2022optical}; in astronomy, multi-band observations of transients enhance temporal and spectral fidelity \cite{kasen2017origin}; and in quantum imaging and optical communications, greater spectral diversity directly increases information capacity \cite{moreau2019imaging}.

Conventional refractive optics, however, are fundamentally limited by material dispersion that shifts the focal plane with wavelength. Practical systems therefore rely on multi-element assemblies optimized for narrow bands, for example, smartphone cameras cover only 400–700 nm (fractional bandwidth $\Delta\lambda/\lambda \approx 0.55$), yet require six or more lenses to achieve acceptable performance \cite{he2022design}. Prior work has sought to extend transmissive imaging across multiple infrared bands using diffractive or hybrid elements, but these approaches remain fundamentally bandwidth-limited. Diffractive kinoform surfaces have enabled dual-band MWIR–LWIR correction, albeit with strong dependence on multi-order diffraction and only modest chromatic control\cite{stone1988hybrid}. Hybrid refractive–diffractive architectures can partially cancel chromatic focal shifts and have extended operation to fractional bandwidths approaching unity \cite{hu2022design, zhang2019design, lidwell1996diffractive}, but typically at the expense of many discrete elements and increased design and manufacturing complexity. Alternative strategies include computationally assisted flat optics,\cite{froch2025beating} which recently demonstrated a 1-cm-aperture, f/2 meta-optic that merges a diffractive phase profile with a learned reconstruction network to deliver imaging across 450–650 nm ($\Delta\lambda/\lambda\approx0.36$). 

\begin{figure}[htb!]
	\centering
	\includegraphics[width=\textwidth]{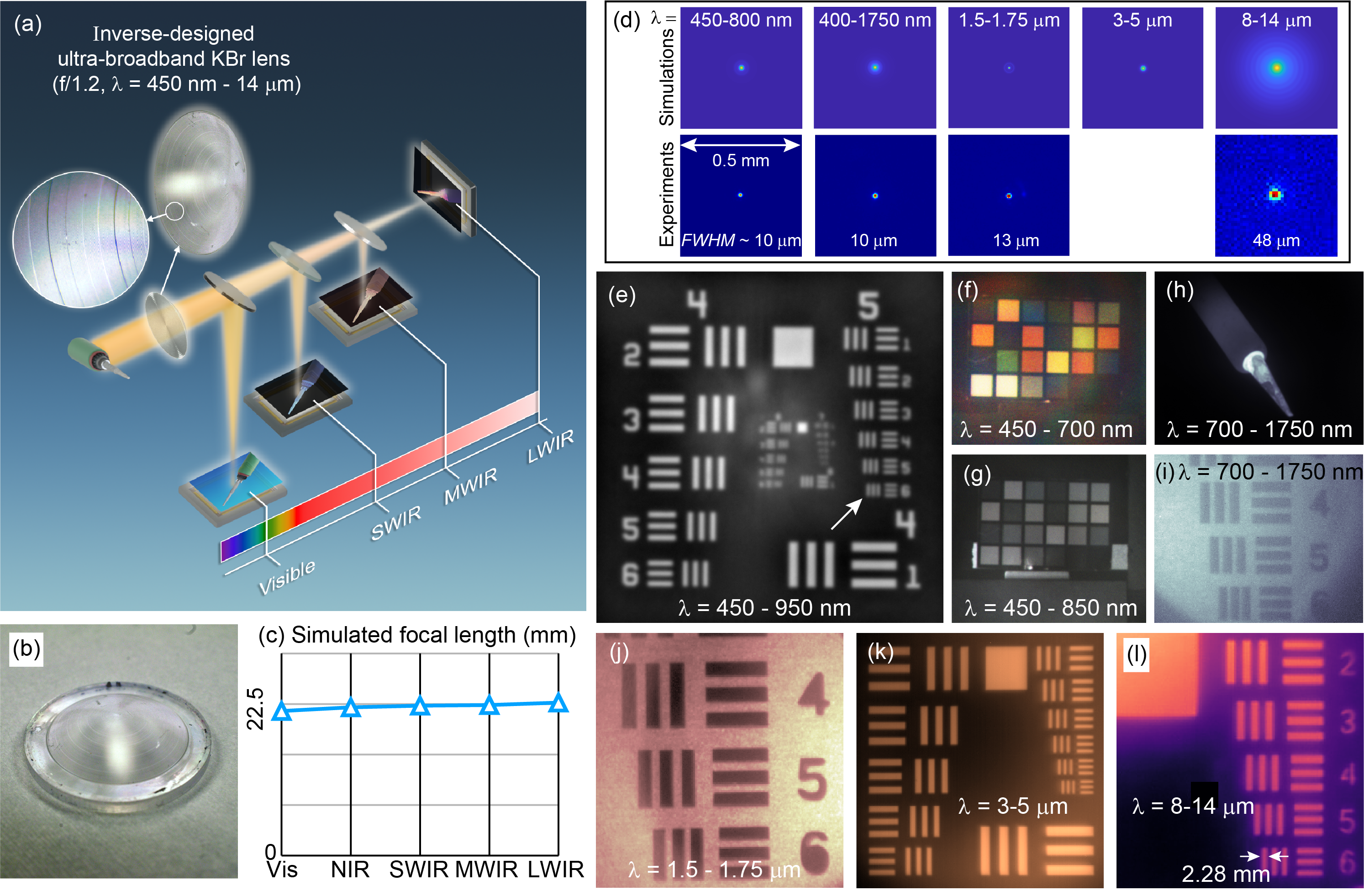} 
	\caption{\textbf{Focusing and imaging from the visible (Vis) to the long-wave infrared using a single lens.} (a) A 19-mm-diameter, single-point diamond-turned potassium bromide (KBr) ultra-broadband lens (nominal focal length 22.5 mm) images the same scene seamlessly from the visible (Vis) to the long-wave infrared (LWIR). The inset shows a micrograph of the microstructured surface, and (b) a photograph of the lens. (c) Simulated focal length, extracted from the peak of the on-axis point-spread function (PSF), remains nearly constant across the full spectral range. (d) Simulated and measured PSFs in representative bands from Vis to LWIR exhibit close agreement, confirming the lens’s achromatic response. (e–l) Representative images acquired in each band demonstrate diffraction-limited performance with minimal chromatic variation. Detector parameters and experimental details are provided in the Supplementary Information.}
	\label{fig:imaging} 
\end{figure}

In contrast, we use large-scale inverse design to create a monolithic, microstructured lens (19-mm diameter; $\sim$22.5-mm focal length) that delivers quasi-achromatic imaging across an unprecedented spectral range, from 450 nm in the visible to beyond 14 $\mu$m in the long-wave infrared. This six-octave span corresponds to a fractional bandwidth greater than 1.9, approaching the theoretical limit of 2. The design discretizes the aperture into concentric 20-$\mu$m zones whose heights are continuously optimized to counteract wavelength-dependent phase delays, thereby suppressing longitudinal chromatic aberration over the full operating range. We used 273 wavelength samples with 50nm spacing for inverse design. As a result, the ultra-broadband (UBB) lens maintains a nearly constant focal length and markedly reduces chromatic focal drift relative to a conventional plano-convex lens (Supplementary Information). This represents more than an order-of-magnitude increase in usable spectral bandwidth compared with state-of-the-art transmissive optics, enabling true VIS–SWIR–MWIR–LWIR imaging using a single passive element, see Fig.~\ref{fig:imaging}(a).

This capability emerges from co-engineering the positive material dispersion of a bulk refractive substrate with the negative structural dispersion of a diffractive microrelief on the same surface, allowing minimizing the longitudinal chromatic aberrations across a large wavelength range. Although earlier hybrid diffractive–refractive approaches have exploited this principle, they have done so over far narrower bandwidths and have required multiple discrete optical elements\cite{stone1988hybrid, davidson1993analytic}. Other demonstrations have integrated shallow diffractive features onto refractive substrates, but only at sub-millimeter apertures\cite{richards2023hybrid} or with correction limited to a few target wavelengths\cite{lin2011study}. To our knowledge, no previous design has achieved continuous, multi-octave quasi-achromatic focusing using a single, macroscopic transmissive optic.

The lens is machined from potassium bromide (KBr), a crystal that remains transparent from the Vis to the LWIR, and therefore enables a single transmissive element across the entire spectral range. KBr’s cubic symmetry and mechanical softness allow nanometer-scale surface precision under diamond turning, yielding optical-quality finishes across the full 19-mm aperture (see Supplement). A conformal parylene coating applied immediately after machining protects the hygroscopic crystal from moisture. Photographs and micrographs of the finished device are shown in Figs.~\ref{fig:imaging}(a)–(b). The resulting 3-mm-thick component is compact, mechanically robust, and bypasses long-standing efficiency–bandwidth limits. Although inverse-designed diffractive lenses can provide chromatic correction, their thin multi-level profiles restrict the available optical path length and reduce focusing efficiency, as recently clarified by Miller\cite{miller2023optics}. Broadband imaging requires sufficient thickness to support many independent spatial channels, a form of overlapping nonlocality that demands at least millimeter-scale propagation depth. Flat metasurfaces, confined to a subwavelength layer, lack this internal volume and cannot sustain the channel count needed for multi-octave achromatization. Our UBB lens, by contrast, uses its extended optical path to achieve practical quasi-achromatic performance across the full VIS–SWIR–MWIR–LWIR range.

We numerically modeled the point-spread function (PSF) across discrete spectral bands—visible (VIS), near-infrared (NIR), short-wave infrared (SWIR), mid-wave infrared (MWIR), and long-wave infrared (LWIR)—to quantify chromatic performance. The focal length at each wavelength was determined from the axial intensity peak, revealing negligible longitudinal chromatic aberration (Fig.~\ref{fig:imaging}c). Simulated (top row) and measured (bottom row) transverse PSFs for representative bands are shown in Fig.~\ref{fig:imaging}d. Good focusing and excellent agreement between simulations and experiments are observed. Experimental PSFs were characterized under band-specific illumination using detectors matched to the corresponding spectral ranges: a silicon CMOS sensor (DMM 27UP031-ML, Imaging Source) for VIS–NIR (450-800 nm), an InGaAs sensor (Sony IMX990, Lucid Vision) for VIS–SWIR (400-1750 nm), a thermoelectrically cooled InSb detector (FLIR SC6000) for MWIR (3 - 5 $\mu$m), and an uncooled microbolometer (FLIR Tau 2 or Boson+) for LWIR (8-14 $\mu$m). In all cases, the distance from the lens to the sensor was kept approximately fixed to 22.5 mm. The imaging performance was experimentally characterized across multiple spectral bands using detectors matched to their respective wavelength ranges (Figs.~\ref{fig:imaging}d-k). Notably, a single lens was employed for all measurements, yet the focal length remained effectively constant from the visible through the long-wave infrared (additional details in Supplement). Each image is labeled with the corresponding wavelength band. Additional experiments detailed in the Supplement confirmed that the lens provides a field-of-view of $\sim32^{\circ}$ and an extended depth-of-field, producing sharp LWIR images of a soldering iron as the object distance varied from 171 mm to beyond 270 mm without adjusting the image plane.

\begin{figure} 
	\centering
	\includegraphics[width=0.75\textwidth]{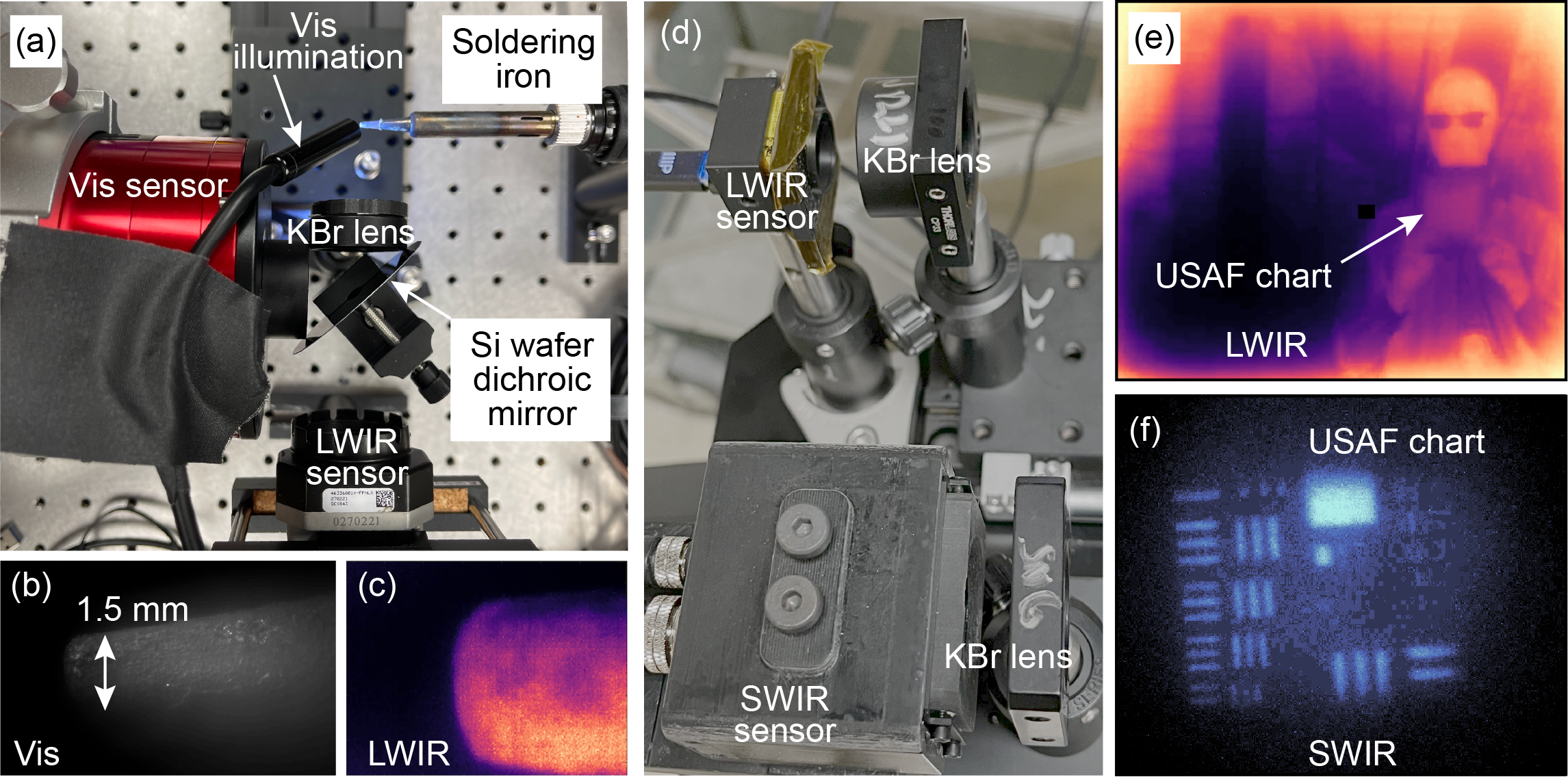}
	\caption{\textbf{Simultaneous two-band imaging.} (a) The UBB lens images a heated soldering iron simultaneously in the visible and LWIR bands. A double-side-polished silicon wafer serves as a dichroic beamsplitter, reflecting visible light toward the CMOS sensor and transmitting thermal radiation to the LWIR detector, both positioned at identical image distances. The resulting co-registered (b) Vis and (c) LWIR images confirm diffraction-limited fidelity and spatial alignment across more than three octaves of the electromagnetic spectrum. See Supplementary Video 1 for combined Vis-LWIR video imaging. (d) Simultaneous SWIR-LWIR imaging with two identical KBr lenses of the same Air Force (AF) resolution target placed about 206 cm away from both cameras. (e) LWIR image with a person holding the AF target and (f) the SWIR image of the same target, showing much higher resolution. See Supplementary Video 2 for combined SWIR-LWIR video imaging.}
	\label{fig:dual_band} 
\end{figure}

 We next demonstrated broadband functionality of the UBB lens by performing simultaneous dual-band imaging spanning the visible and long-wave infrared (LWIR) regimes (Fig.~\ref{fig:dual_band}a). A soldering iron heated to $\approx400^{\circ}$C served as the object, illuminated by a broadband white LED. A 2-inch double-side-polished silicon wafer acted as a beamsplitter, reflecting the visible component toward a monochrome CMOS detector (ASI2600MM Pro, ZWO) while transmitting thermal emission to an uncooled LWIR camera (Tau 2, FLIR). The object and image distances were fixed at u $\approx$ 32 mm and v $\approx$ 52.5 mm for both channels. The resulting images exhibit high clarity in both visible (after post-processing, Fig.~\ref{fig:dual_band}b) and LWIR (no post-processing, Fig.~\ref{fig:dual_band}c) bands. Simultaneous video recordings from the visible and LWIR detectors confirmed broadband image registration and consistent performance across both bands (see Supplementary Video 1). When the soldering iron was unheated, it appeared only in the visible channel; as it reached operating temperature, a distinct LWIR image emerged, validating co-registered imaging of visible and thermal fields. Translating the hot tip across the field of view produced matched motion in both modalities, underscoring the achromatic alignment of the KBr lens. Representative frames from the synchronized recordings are shown in Figs. S2.10b–d.

 In a separate experiment, we used two identical KBr lenses paired with side-by-side cameras to record SWIR and LWIR images, simultaneously (Fig. \ref{fig:dual_band}d). A resolution chart back-illuminated by a SWIR LED array was placed approximately 206 cm from the lenses. The SWIR channel captured the fine features of the chart with high fidelity, while the LWIR channel revealed the thermal emission from the person holding it (Figs. \ref{fig:dual_band}e,f; Supplementary Video 2). Together, the co-registered SWIR and LWIR videos demonstrate true dual-band imaging within a single optical architecture and highlight the ability of the KBr lens to operate effectively across widely separated infrared regimes.

\begin{figure} 
	\centering
	\includegraphics[width=\textwidth]{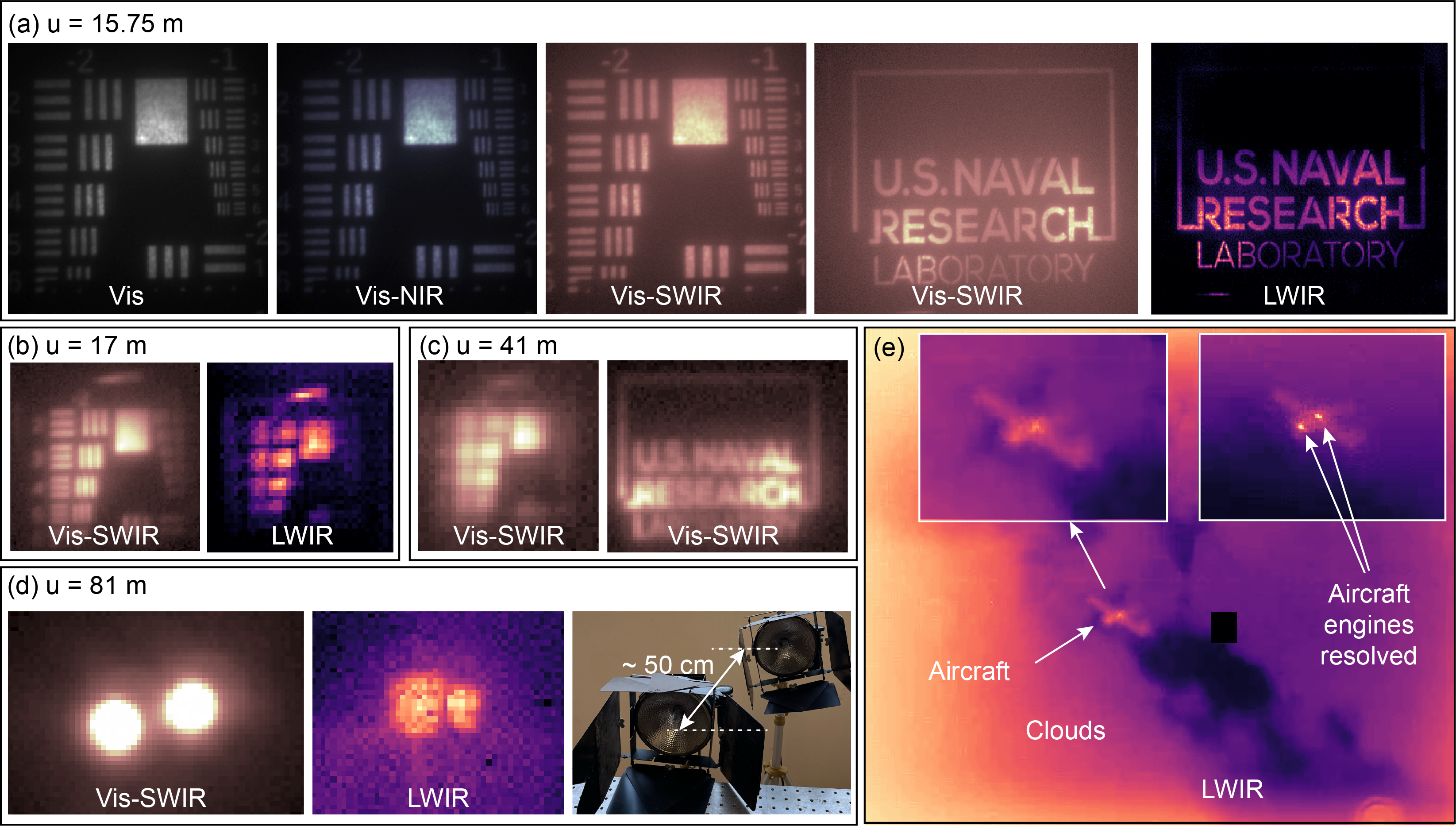}
	\caption{\textbf{Broadband long-range imaging across visible to LWIR bands.} Long-range, multi-spectral imaging was performed using the dual-camera configuration shown in Fig.~\ref{fig:dual_band}d. Test objects—including an AF resolution chart and an NRL logo—were back-illuminated by natural sunlight or halogen lamps. (a) Imaging of the AF target at a distance of 15.75 m, captured in (left to right) the visible–SWIR, visible, and NIR bands, and the NRL logo in the visible–SWIR and LWIR bands. (b) AF target imaged at 17 m in the visible–SWIR and LWIR bands. (c) NRL logo and AF target imaged simultaneously in the visible–SWIR band at 41 m. (d) Two halogen lamps (inset photograph) acting as quasi-point sources at 81 m are distinctly resolved in both the visible–SWIR (left) and LWIR (right) bands. (e) Frame from Supplementary Video 3 showing LWIR imaging of an aircraft from the NRL rooftop, where the two hot engines are clearly resolved.}
	\label{fig:long_range_imaging} 
\end{figure}

We next evaluated the UBB lens for long-range, multi-band imaging using the dual-camera configuration in Fig. \ref{fig:dual_band}d. Targets were placed at distances ranging from $\sim15$ m to 81 m. As summarized in Fig. \ref{fig:long_range_imaging}, the system produced high-quality images of the AF resolution chart and the NRL logo across the Vis, NIR, VIs-to-SWIR, and LWIR bands. At a range of 81 m, two halogen lamps separated by $\sim50$ cm were clearly resolved in both the Vis–SWIR and LWIR channels, corresponding to an angular resolution of about 6 milliradians (Fig. \ref{fig:long_range_imaging}d). Finally, we deployed the dual-camera system on the roof of a building at NRL to image aircraft departing from a nearby airport. An example LWIR frame (Fig. \ref{fig:long_range_imaging}e, Supplementary Video 3) distinctly resolves the two hot engines of an aircraft estimated to be $>2$ km from the lens. Also see additional video of vehicles in a parking lot (Supplementary Video 4).  
 
\begin{figure} 
	\centering
	\includegraphics[width=\textwidth]{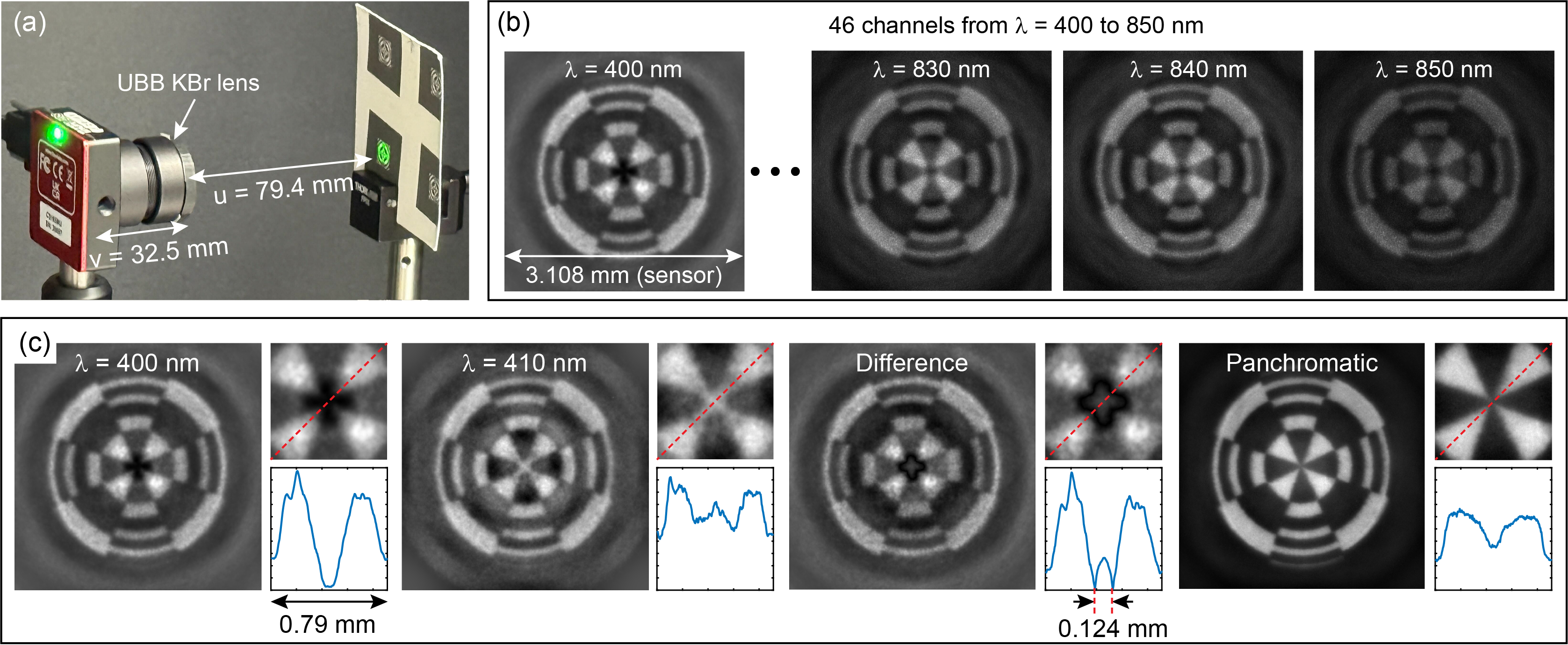}
	\caption{\textbf{Hyperspectral and super-resolution imaging across the visible–near-infrared spectrum.} A reflective target was illuminated with narrowband light from a tunable supercontinuum laser, and images were recorded through the single KBr lens without refocusing across wavelengths (constant object and image distances). (a) Photograph of the experimental setup. (b) Representative hyperspectral images selected from 46 wavelength channels spanning $400–850$ nm. (c) Demonstration of super-resolution capability: the difference between the 400 nm and 410 nm images reveals a central feature $<$ 0.2 mm wide, corresponding to an effective angular resolution of $\approx$ 1.6 mrad.}
	\label{fig:Vis_HSI} 
\end{figure}

We next assessed the UBB lens for hyperspectral imaging across the VIS–NIR band (400 to 850 nm) using an Imatest Reflective SFRplus chart (Fig. \ref{fig:Vis_HSI}a). The experiment was conducted in reflection using spectrally tunable laser illumination, with the center wavelength stepped in 10-nm increments (and bandwidth = 10 nm). After an initial focus adjustment at 550 nm, the focal setting remained fixed for all subsequent wavelengths. Representative raw and processed images are shown in Figs. S2.7e–f, and the full acquisition and calibration procedures are detailed in the Supplementary Information (Figs. S2.7a–b; Tables S2.2–S2.3). Across all 46 measured spectral channels, the lens maintains uniform sharpness and contrast (Fig. \ref{fig:Vis_HSI}b), confirming achromatic performance over more than an optical octave. The hyperspectral dataset also reveals how wavelength diversity can enhance spatial discrimination. As an illustration, Fig. \ref{fig:Vis_HSI}c compares images acquired at 400 nm and 410 nm. Subtle differences in the central region become pronounced when the two images are differenced, whereas the same fine features are blurred in the panchromatic (wavelength-averaged) image. Insets show magnified views and corresponding line scans, highlighting that the difference image resolves structures that are otherwise unresolved. Specifically, we resolve a feature of width $\sim 0.124$ mm from a distance of 79.4 mm, corresponding to an angular resolution of 1.6 mrads. This demonstration of spectral information enabling spatial super-resolution is achieved here with a single-aperture UBB lens, suggesting that such information-rich imaging can be extended seamlessly across far broader wavelength ranges.

\begin{figure} 
	\centering
	\includegraphics[width=\textwidth]{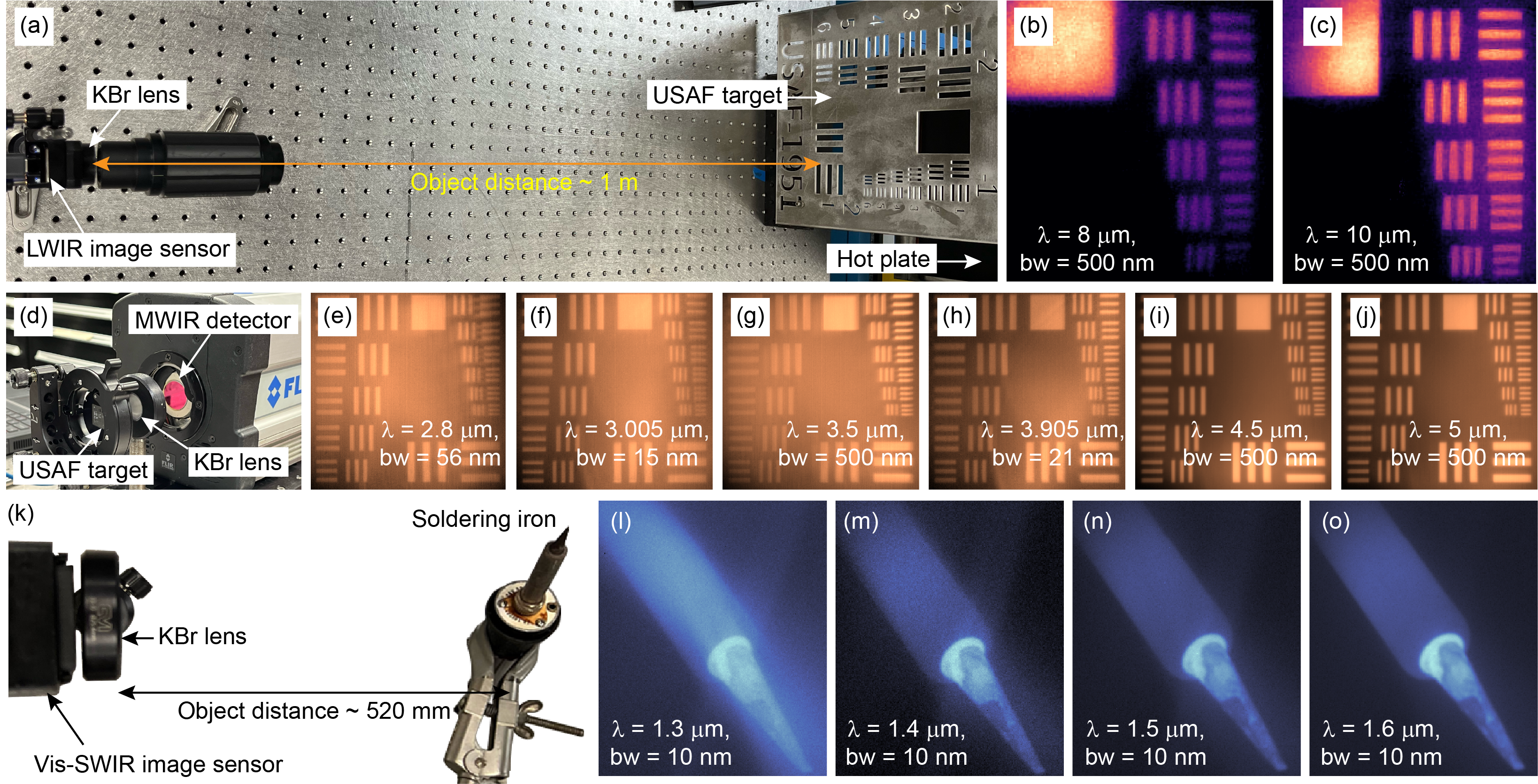}
	\caption{\textbf{Broadband hyperspectral imaging across SWIR–MWIR–LWIR bands using a single KBr lens.} A single KBr lens was used to perform hyperspectral imaging across the short-, mid-, and long-wave infrared regimes. (a) Experimental setup for LWIR imaging of an AF resolution target back-illuminated by a calibrated hot plate. Wavelength-resolved images acquired with narrowband filters centered at (b) 8 $\mu$m and (c) 10 $\mu$m (bandwidth = 500 nm) exhibit high sharpness and contrast. (d) Setup for MWIR hyperspectral imaging of the same target. Corresponding images were captured at (e) 2.8 $\mu$m (bandwidth = 56 nm), (f) 3.01 $\mu$m (15 nm), (g) 3.5 $\mu$m (500 nm), (h) 3.91 $\mu$m (21 nm), (i) 4.5 $\mu$m (500 nm), and (j) 5.0 $\mu$m (500 nm). (k) Setup for SWIR hyperspectral imaging of a heated soldering iron, with images acquired at center wavelengths of (l) 1.6 $\mu$m, (m) 1.5 $\mu$m, (n) 1.4 $\mu$m, and (o) 1.3 $\mu$m (bandwidth = 10 nm for all cases). Across all measurements, the same KBr lens produced diffraction-limited performance and maintained consistent focus from 1.3 $\mu$m to 10 $\mu$m.}
	\label{fig:IR_HSI} 
\end{figure}
We further demonstrate broadband hyperspectral imaging using the same UBB lens across the SWIR, MWIR, and LWIR bands (Fig.~\ref{fig:IR_HSI}). Photographs of the LWIR, MWIR, and SWIR experimental configurations are shown in Figs.~\ref{fig:IR_HSI}a, d, and k, respectively. In all cases, wavelength-resolved images were acquired using narrowband interference filters. For LWIR imaging, a large-area AF resolution chart back-illuminated by a calibrated blackbody source (Mikron M345 Blackbody, 12' $\times$ 12') produced well-resolved features at 8 $\mu$m and 9 $\mu$m (0.5 $\mu$m bandwidth; Figs.\ref{fig:IR_HSI}b–c). MWIR measurements used a smaller version of the chart, again back-illuminated by a blackbody, yielding sharp images at 2.8 $\mu$m, 3.005 $\mu$m, 3.5 $\mu$m, 3.905 $\mu$m, 4.5 $\mu$m, and 5 $\mu$m (Figs.~\ref{fig:IR_HSI}e–j). Notably, the 3.005 $\mu$m channel, with a bandwidth of only 15 nm, exhibits high contrast and signal-to-noise ratio, underscoring the lens’s ability to maintain performance even under extremely narrow spectral isolation. To probe SWIR performance, we imaged a soldering iron using a VIS–SWIR sensor without auxiliary illumination. Narrowband images (10 nm bandwidth) at 1.6 $\mu$m, 1.5 $\mu$m, 1.4 $\mu$m, and 1.3 $\mu$m (Figs.~\ref{fig:IR_HSI}l–o) retain high fidelity and fine spatial detail despite the low photon flux at these wavelengths, all without any post-processing. Altogether, the UBB lens enabled hyperspectral imaging over \textbf{58 discrete channels}: 46 in the VIS–NIR, 4 in the SWIR, 6 in the MWIR, and 2 in the LWIR, spanning more than three octaves of optical bandwidth. To our knowledge, this constitutes the first demonstration of hyperspectral imaging from the visible through the long-wave infrared using a single aperture.

Near-infrared (NIR) light penetrates scattering media such as haze, fog, or tissue \cite{ma2021deep} more effectively than visible light, producing sharper, higher-contrast images over long distances. Conventional systems acquire visible and NIR images separately, using distinct cameras, filters, or lenses, and then merge them computationally to recover scene detail. In contrast, our KBr lens inherently forms broadband (Vis–NIR) images in a single capture, eliminating the need for image fusion. To assess its performance through scattering media, we introduced a diffusive layer by stacking 20 transparency sheets (901 Clear Film, Highland) between the lens–sensor assembly (ASI2600MM Pro) and a test scene illuminated by a broadband halogen source (Fig.~\ref{fig:scatt}a, See Supplement for details). The scene comprised a printed resolution chart, a miniature color checker, and small objects, with object and image distances of 760 mm and 21 mm, respectively. Sequential imaging with red, green, and blue filters yielded the RGB composite (Fig.~\ref{fig:scatt}b), in which most fine details were obscured by scattering. When imaged without any filter, capturing the full visible–NIR spectrum, the resulting image (Fig.~\ref{fig:scatt}c) revealed markedly sharper features and enhanced contrast. Averaged line profiles through the white-dashed region confirm a substantial contrast improvement for broadband (Vis–NIR) imaging compared to visible-only captures, demonstrating the advantage of simultaneous multispectral transmission through the scattering medium. 

To further demonstrate the potential of "seeing" through obscurants, we used the KBr lens with the LWIR sensor to image scenes containing objects that were obscured in the visible band, but not in the LWIR. A hand wearing a blue nitrile glove and concealed by a white plastic sheet (Fig.~\ref{fig:scatt}d, See Supplement for additional details), both transparent to LWIR but opaque in the visible, was invisible to the eye, yet clearly resolved in thermal images (Fig.~\ref{fig:scatt}e). When the hand was covered by two layers of camouflage fabric, which effectively conceal it in the visible band (Fig.~\ref{fig:scatt}f), the LWIR images still revealed its thermal signature (Fig.~\ref{fig:scatt}g; object and image distances: 864 mm and 22.9 mm). Together, these experiments underscore the complementary nature of visible and LWIR imaging and highlight the ability of LWIR optics to reveal features hidden from view in conventional visible imaging.
\begin{figure} 
	\centering
	\includegraphics[width=\textwidth]{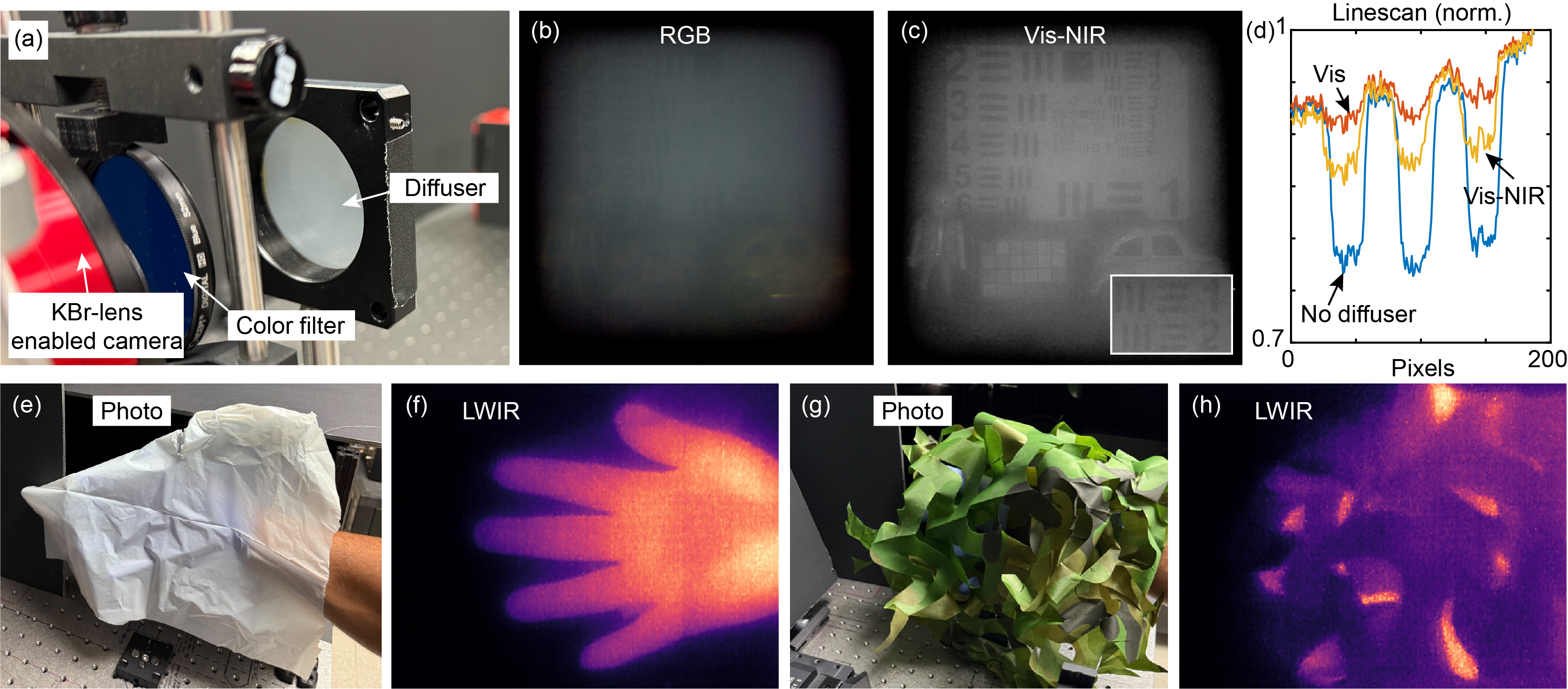}
	\caption{\textbf{Broadband spectral imaging with the KBr lens reveals hidden features through scattering and obscuring media.} (a–d) Imaging through a strongly scattering diffuser. (a) Photograph of the diffuser. (b) RGB image showing severe loss of contrast and detail. (c) Broadband Vis–NIR image captured with the KBr lens, exhibiting markedly improved sharpness and contrast. (d) Averaged line profiles through the white dashed region highlight the enhanced contrast compared to the RGB and no-diffuser control cases. (e–h) LWIR imaging through visible obscurants. A gloved hand concealed beneath a white plastic sheet or camouflage fabric is invisible in visible photographs (e,g) but clearly resolved in corresponding LWIR images (f,h), where thermal emission from the hand is revealed.
}
	\label{fig:scatt} 
\end{figure}

\begin{figure} 
	\centering
	\includegraphics[width=\textwidth]{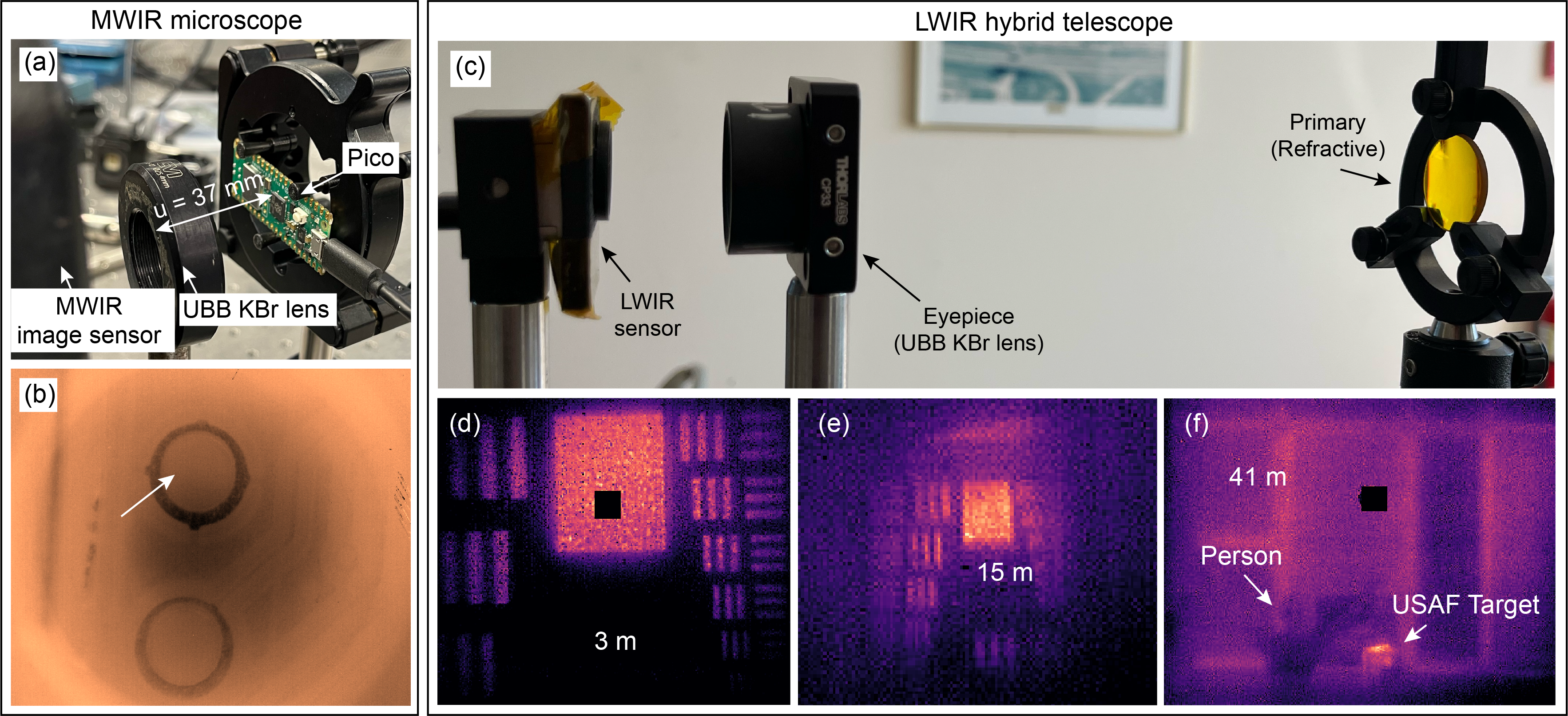}
	\caption{\textbf{Application of the UBB lens in a MWIR microscope and a hybrid LWIR telescope.} (a) Photograph of the MWIR microscope, which uses the KBr UBB lens as its objective to image a Raspberry Pi Pico circuit board. (b) Representative frame from Supplementary Video 5, revealing two capacitors and multiple localized hot spots—including current-carrying traces. (c) Photograph of the hybrid LWIR telescope in which a conventional ZnSe refractive lens acts as the primary objective and the KBr UBB lens serves as the eyepiece. This configuration enables broadband imaging across long standoff distances. LWIR images of an AF resolution target back-illuminated by a heat lamp are shown at (d) 3 m and (e) 15 m, demonstrating sustained sharpness and contrast across more than a five-fold change in range. (f) LWIR image acquired at 41 m using a second hybrid telescope in which the ZnSe primary is replaced with a compound refractive objective; additional details are provided in the Supplementary Information.
}
	\label{fig:Muscope_hybrid} 
\end{figure}

To highlight the adaptability of the UBB lens across infrared modalities, we demonstrated an MWIR microscope by positioning the lens 37 mm above an actively powered Raspberry Pi Pico (Fig.~\ref{fig:Muscope_hybrid}a). A representative frame from Supplementary Video 5 (Fig.~\ref{fig:Muscope_hybrid}b) reveals sub-millimeter components and thermally active traces with high contrast, demonstrating the lens’s ability to resolve fine features solely from their MWIR emission. 

In a separate experiment, we deployed the UBB lens as the eyepiece in a hybrid Keplerian telescope, coupled to a conventional refractive primary (Fig.~\ref{fig:Muscope_hybrid}c). This configuration enabled high-fidelity LWIR imaging of an AF resolution target back-illuminated by a heat lamp across standoff distances spanning 3 m to 41 m (Figs.~\ref{fig:Muscope_hybrid}d–f). To demonstrate the robustness and modularity of the approach, we repeated the measurements using different refractive primaries, each yielding consistent image sharpness and contrast. Together, these results establish that the UBB lens can reliably augment conventional refractive optics, extending their performance in broadband infrared imaging systems.

We show that a single monolithic transmissive lens can image continuously from the Vis to the LWIR, a $>14\thinspace\mu$m span corresponding to a fractional bandwidth $>1.9$. By combining inverse design with precision diamond turning of KBr, we realized a 19-mm-diameter, 22.5-mm-focal-length optic whose focal position remains nearly constant across six optical octaves. This level of chromatic suppression far exceeds that of conventional refractive, diffractive, or hybrid systems, which typically require multi-element assemblies to correct even a single octave. Demonstrations of ultra-broadband focusing, 58-channel hyperspectral imaging, and stable focal length from 450 nm to beyond 14 $\mu$m establish a new regime of quasi-achromatic transmissive optics.

The same lens supports imaging modes that normally require separate optical trains. It enables simultaneous visible/LWIR and SWIR/LWIR imaging, long-range multispectral imaging beyond 80 m, and VIS–NIR hyperspectral imaging without refocusing. Its robustness to haze, fabrics, and other obscurants highlights the benefits of NIR and LWIR transparency. When paired with a conventional refractive primary, the lens also forms a compact hybrid telescope with tunable magnification.

Conceptually, this work introduces a new strategy for extreme bandwidth: dispersive phase engineering through locally optimized surface microrelief. By co-optimizing micrometer-scale zones across dozens of wavelengths, inverse design bypasses long-standing trade-offs among bandwidth, element count, and manufacturability. The result is a scalable monolithic optic that replaces complex multi-element assemblies. The ability to collect visible, thermal, and spectral information through one compact element opens opportunities in planetary exploration, environmental monitoring, autonomous systems, surveillance, and biomedical imaging. More broadly, combining inverse design with broadband-transparent materials points toward compact, information-rich imaging systems whose spectral reach is no longer limited by traditional optical constraints. 


\clearpage 

%
\bibliography{science_template} 
\bibliographystyle{sciencemag}


\section*{Acknowledgments}
We thank Tim Olsen (Omega Optical) for expert support with diamond turning of the KBr optics. We are grateful to Paul Ricketts for lending us the ZWO ASI2600MM Pro sensor. We acknowledge insightful discussions with Alex Kormos, Bob Owen, and Paul Leonelli on LWIR testing. We also acknowledge Alec Ikei for assistance with experiments conducted at the Naval Research Laboratory.

\paragraph*{Funding:}
Office of Naval Research (ONR) \#N000142512122 and \#N000142212014. 
National Science Foundation (NSF) \#2229036. 
\paragraph*{Author contributions:}
NSQ: Performed IR experiments and metrology of the lens. 
AM: Performed Vis-NIR and LWIR experiments. 
JDH: Performed imaging and PSF experiments. 
NB: Performed inverse design and all simulations. 
FS: Performed IR experiments.
RM: Conceptualized and managed the project, performed experiments, analyzed data and wrote the manuscript. 
All authors edited the manuscript. 
\paragraph*{Competing interests:}
RM and NB have financial interest in Oblate Optics, which is commercializing the technology herein.
\paragraph*{Data and materials availability:}
All materials are available in the manuscript and the supplement. Any other data is available from the corresponding author upon reasonable request. 

\end{document}